\documentstyle[prd,aps,floats,tighten,epsfig,graphicx]{revtex}

\newcommand{\beq}{\begin{equation}}
\newcommand{\eeq}{\end{equation}}

\newcommand{\etal}{{\rm et al.}}

\newcommand{\dams}{\Delta a_\mu({\rm SUSY})}

\newcommand{\app}[3]{Astropart.\ Phys.\ {\bf #1}, #3 (#2)}

\newcommand{\hepph}[1]{{\tt hep-ph/#1}}

\newcommand{\prep}[3]{Phys.\ Rep.\ {\bf #1}, #3 (#2)}
\newcommand{\plb}[3]{Phys.\ Lett.\ B\ {\bf #1}, #3 (#2)}
\newcommand{\npb}[3]{Nucl.\ Phys.\ B\ {\bf #1}, #3 (#2)}
\newcommand{\cpc}[3]{Comm.\ Phys.\ Comm.\ {\bf #1}, #3 (#2)}
\renewcommand{\apj}[3]{Astrophys.\ J.\ {\bf #1}, #3 (#2)}

\renewcommand{\prl}[3]{Phys.\ Rev.\ Lett. {\bf #1}, #3 (#2)}
\renewcommand{\prd}[3]{Phys.\ Rev.\ D\ {\bf #1}, #3 (#2)}
\renewcommand{\rmp}[3]{Rev.\ Mod.\ Phys.\ {\bf #1}, #3 (#2)}

\begin{document}
\draft


\title{Implications of muon anomalous magnetic moment for
supersymmetric dark matter}

\author{E.~A.~Baltz}
\address{ISCAP, Columbia Astrophysics Laboratory, 550 W 120th St., Mail Code
5247, New York, NY 10027}
\author{P.~Gondolo} \address{Department of Physics, Case Western Reserve
University, 10900 Euclid Ave., Cleveland, OH 44106-7079}

\twocolumn[\hsize\textwidth\columnwidth\hsize\csname@twocolumnfalse\endcsname
\maketitle


\begin{abstract}
The anomalous magnetic moment of the muon has recently been measured to be in
conflict with the Standard Model prediction with an excess of $2.6\sigma$.
Taking the excess at face value as a measurement of the supersymmetric
contribution, we find that at 95\% confidence level it imposes an upper bound
of 500 GeV on the neutralino mass and forbids higgsinos as being the bulk of
cold dark matter.  Other implications for the astrophysical detection of
neutralinos include: an accessible minimum direct detection rate, lower bounds
on the indirect detection rate of neutrinos from the Sun and the Earth, and a
suppression of the intensity of gamma ray lines from neutralino annihilations
in the galactic halo.
\end{abstract}


\vskip 2pc]\narrowtext

Recently, the Brookhaven AGS experiment 821 measured the anomalous magnetic
moment of the muon $a_\mu=(g-2)/2$ with three times higher accuracy than it was
previously known \cite{DATA}. Their result is higher than the Standard Model
prediction at greater than $2.6\sigma$.  One well-known possibility is that
supersymmetric corrections to $a_\mu$ are responsible for this discrepancy
\cite{a_mu_old,moroi,SM}.  In this Letter, we take the approach that all the
measured discrepancy is due to supersymmetric contributions, and discuss the
implications of this measurement for searches of neutralino dark matter.

There are two caveats to our approach.  The first is that there is some
disagreement on what the Standard Model prediction is, primarily in the
hadronic contribution.  There remain theoretical evaluations for which the new
experimental result agrees with the Standard Model \cite{nodiscrepancy}.  The
second caveat is that supersymmetry is only one possibility for physics beyond
the Standard Model that could contribute to $a_\mu$.  Other possibilities
include (but are not limited to) radiative fermion masses, extended technicolor
and anomalous gauge boson couplings, as summarized in Ref.~\cite{SM}.

The lightest stable supersymmetric particle in the Minimal Supersymmetric
Standard Model (MSSM) is most often the lightest of the neutralinos, which are
superpositions of the superpartners of the neutral gauge and Higgs bosons,
\begin{equation}
\tilde{\chi}^0_1 =
N_{11} \tilde{B} + N_{12} \tilde{W}^3 +
N_{13} \tilde{H}^0_1 + N_{14} \tilde{H}^0_2.
\end{equation}

For many values of the MSSM parameter space, the relic density $\Omega_\chi
h^2$ of the (lightest) neutralino is of the right order of magnitude for the
neutralino to constitute at least a part, if not all, of the dark matter in the
Universe (for a review see Ref.~\cite{jkg96}). Here $\Omega_{\chi}$ is the
density in units of the critical density and $h$ is the present Hubble constant
in units of $100$ km s$^{-1}$ Mpc$^{-1}$. Present observations favor $h =
0.7\pm 0.1$, and a total matter density $\Omega_{M} = 0.3 \pm 0.1$, of which
baryons contribute roughly $\Omega_bh^2\approx0.02$ \cite{cosmparams}.  Thus we
take the range $0.052\le\Omega_\chi h^2\le0.236$ as the cosmologically
interesting region. We are also interested in models where neutralinos are not
the only component of dark matter, so we also separately consider models with
arbitrarily small $\Omega_\chi h^2<0.236$.

We have explored a variation of the MSSM.  Our framework has seven free
parameters: the higgsino mass parameter $\mu$, the gaugino mass parameter
$M_{2}$, the ratio of the Higgs vacuum expectation values $\tan \beta$, the
mass of the $CP$--odd Higgs boson $m_{A}$, the scalar mass parameter $m_{0}$
and the trilinear soft SUSY--breaking parameters $A_{b}$ and $A_{t}$ for third
generation squarks.  Our framework is more general than the supergravity
framework, in that we do not impose radiative electroweak symmetry breaking nor
GUT unification of the scalar masses and trilinear couplings.  The only
constraint from supergravity that we impose is gaugino mass unification, though
the relaxation of this constraint would not significantly alter our results.
We assume that R-parity is conserved, stabilizing the lightest superpartner.
(For a more detailed description of the models we use, see
Refs.~\cite{bg,coann,jephd}.)

\begin{table}
\begin{tabular}{rrrrrrrr}
Parameter & $\mu$ & $M_{2}$ & $\tan \beta$ & $m_{A}$ & $m_{0}$ &
$A_{b}/m_{0}$ & $A_{t}/m_{0}$ \\
Unit & GeV & GeV & 1 & GeV & GeV & 1 & 1 \\ \hline Min & -50000 &
-50000 & 1.0 & 0        & 100 & -3 & -3 \\
Max & 50000 & 50000 & 60.0 & 10000 & 30000 & 3 & 3 \\ 
\end{tabular}
\caption{ The ranges of parameter values used in the MSSM scans of
Refs.~\protect\cite{bg,coann,bub,neutrate,other_db}.  In this Letter, we use
approximately 79,000 models that were not excluded by accelerator constraints
before the recent $a_\mu$ measurement. }
\label{tab:scans}
\end{table}

As a scan in MSSM parameter space, we have used the database of MSSM models
built in Refs.~\cite{bg,coann,bub,neutrate,other_db}. The overall ranges of the
seven MSSM parameters are given in Table~\ref{tab:scans}.  The database
embodies one--loop corrections for the neutralino and chargino masses as given
in Ref.~\cite{neuloop}, and leading log two--loop radiative corrections for the
Higgs boson masses as given in Ref.~\cite{feynhiggs}. The database contains a
table of neutralino--nucleon cross sections and expected detection rates for a
variety of neutralino dark matter searches.

The database also includes the relic density of neutralinos $\Omega_{\chi}
h^2$.  The relic density calculation in the database is based on
Refs.~\cite{coann,GondoloGelmini} and includes resonant annihilations,
threshold effects, finite widths of unstable particles, all two--body
tree--level annihilation channels of neutralinos, and coannihilation processes
between all neutralinos and charginos.

We examined each model in the database to see if it is excluded by the most
recent accelerator constraints. The most important of these are the LEP bounds
\cite{pdg2000} on the lightest chargino mass 
($ m_{\chi_{1}^{+}} > 88.4$ GeV for $| m_{\chi_{1}^{+}} - m_{\chi^{0}_{1}} | >
3 $ GeV and $ m_{\chi_{1}^{+}} >67.7 $ GeV otherwise)
and on the lightest Higgs boson mass $m_{h}$ (which ranges from 91.5--112 GeV
depending on $\tan\beta$) and the constraints from $b \to s \gamma$
\cite{cleo} (we used the LO implementation in DarkSUSY \cite{DarkSUSY}).

The results of Brookhaven AGS experiment E821 \cite{DATA} for the anomalous
magnetic moment of the muon, $a_\mu=(g-2)/2$, compared with the predicted
Standard Model value are
\begin{equation}
a_\mu({\rm exp})-a_\mu({\rm SM})=(43\pm16)\times10^{-10}.
\label{eq:disc}
\end{equation}
This represents an excess of $2.6\sigma$ from the standard model value given in
Ref.~\cite{SM}.

The anomalous magnetic moment $a_\mu$ is quite sensitive to supersymmetry, as
has been calculated by several authors \cite{a_mu_old,moroi,SM}.
Supersymmetric corrections to $a_\mu$, $\dams$, can be either positive or
negative, so in significantly reducing the errors in the measurement of
$a_\mu$, models with negative $\dams$ can be ruled out at high confidence.

We assume that the entire discrepancy (Eq.\ref{eq:disc}) is made up by
supersymmetric corrections, and investigate the implications for the MSSM
parameter space. We consider a 95\% $(2\sigma)$ confidence region for the
supersymmetric contribution, accepting the following range of $\dams$
\begin{equation}
10\times10^{-10}\le\dams\le75\times10^{-10}.
\end{equation}
We compute $\dams$ for the models in the database using the full calculation in
Ref.~\cite{moroi}.

In Fig.~\ref{fig:mxzg} we plot the ratio of gaugino and higgsino fractions
against the mass for the lightest neutralino in a large sample of models. This
ratio is defined as
\begin{equation}
\frac{Z_g}{1-Z_g} = \frac{ |N_{11}|^2 + |N_{12}|^2}{|N_{13}|^2 + |N_{14}|^2}.
\end{equation}
We show the allowed region, with and without the new constraint on $\dams$, in
two cosmological cases.  On the left, we only require that $\Omega_\chi
h^2<0.236 $, whereas on the right, we consider models where the dark matter
could be entirely neutralinos, with the previously mentioned cosmologically
interesting range for $\Omega_\chi h^2$.  In both cases, models allowed before
the $\dams$ constraint are plotted as crosses, and models respecting the
$\dams$ constraint are plotted as crossed circles.

\begin{figure*}[!ht]
\centerline{
\epsfig{file=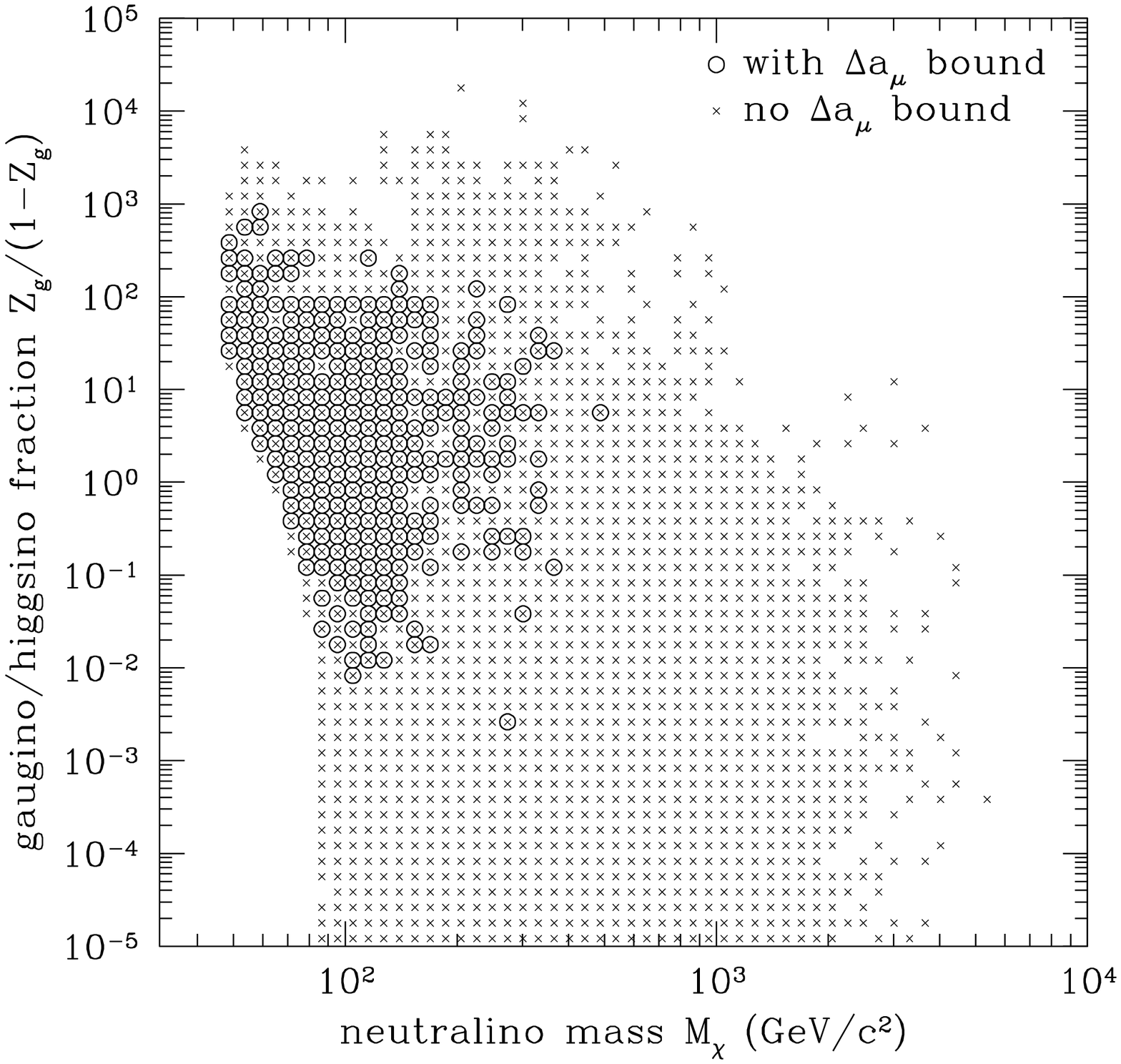,width=0.415\textwidth}
\epsfig{file=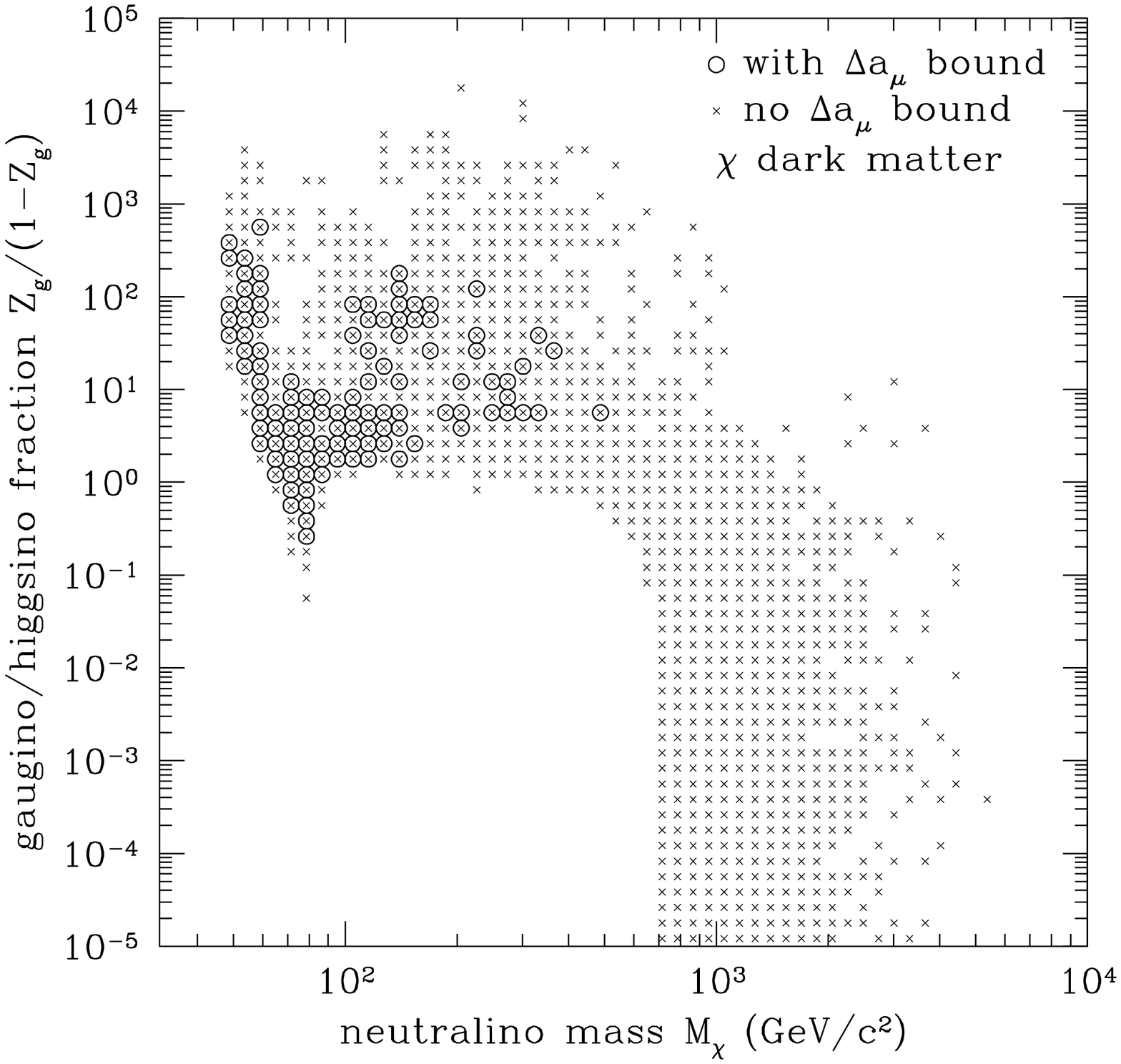,width=0.415\textwidth}}
\caption{Gaugino/higgsino fraction versus mass for the lightest neutralino.  In
the left panel, we plot our set of models allowed by cosmology, but not
requiring that $\Omega_\chi$ be large enough to account for the dark matter.
In the right panel, we apply the constraint that the dark matter is
neutralinos, as discussed in the text.  Crosses indicate previously allowed
models, and the crossed circles indicate models allowed after imposing the
$\dams$ bound.}
\label{fig:mxzg}
\end{figure*}

The~most~pronounced effect of applying the $\dams$ bound is an upper limit of
500 GeV on the neutralino mass. The previous bound of 7 TeV was cosmological,
that is from the constraint $\Omega_\chi h^2<1$ \cite{coann}.  We now find
that the bound from $\dams$ on the neutralino mass is significantly more
stringent.  We note, however, that in taking the $3\sigma$ range of the
experiment, the Standard Model value is included, and the new bound is
completely removed.

Another interesting effect of applying the $\dams$ bound appears when we impose
that the neutralino constitutes the bulk of cold dark matter ($ 0.052 \le
\Omega_\chi h^2 \le 0.236$). In this case, the neutralino must have at least a
10\% admixture of gauginos. Therefore, we can make the claim that neutralino
dark matter can not be very purely higgsino-like. The experimental bound on
$\dams$ disfavors higgsino dark matter even without the theoretical
assumption of supergravity.

We now discuss the implications of these new constraints for astrophysical
searches for neutralino dark matter.

\begin{figure*}[!ht]
\centerline{
\epsfig{file=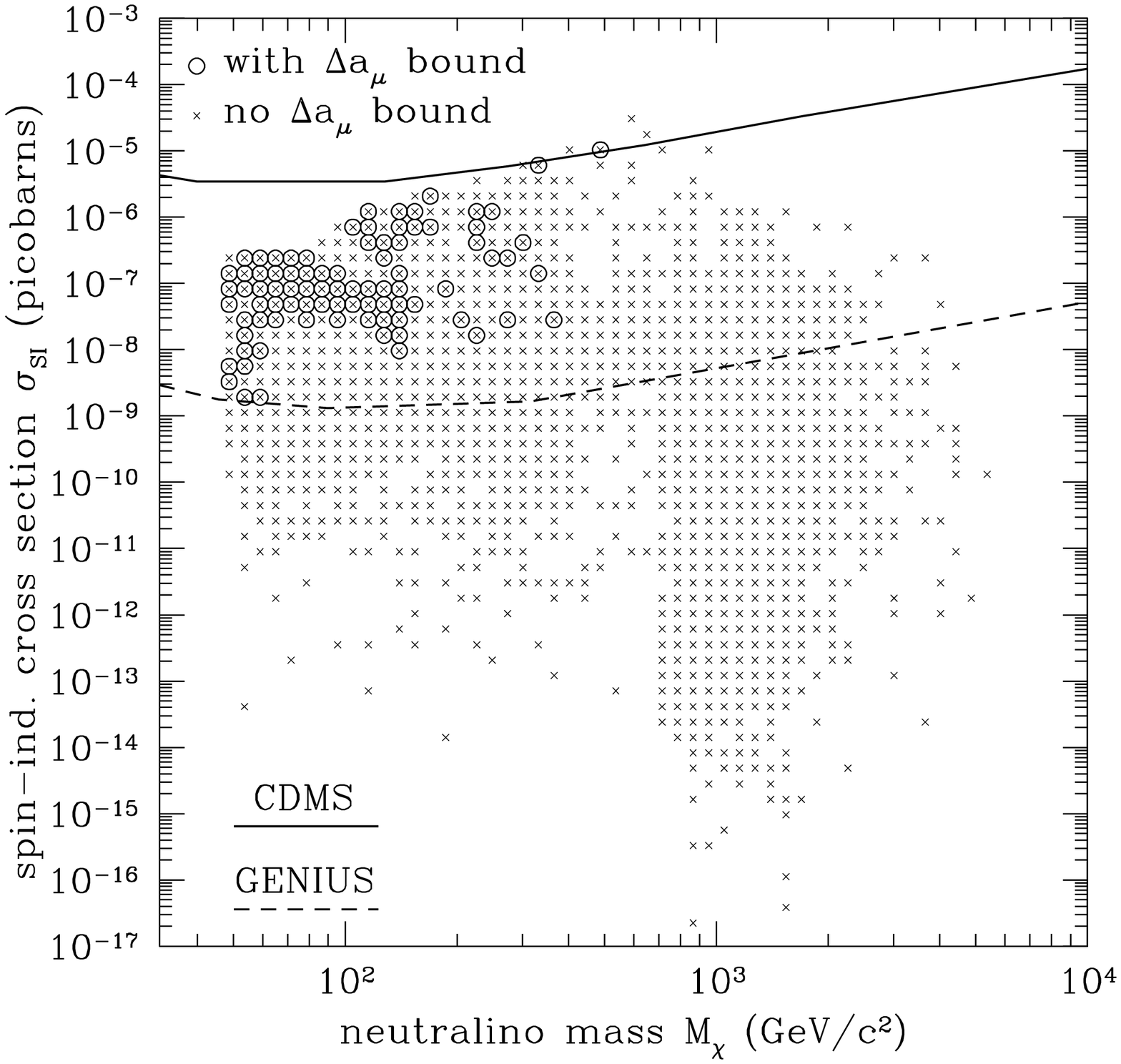,width=0.415\textwidth}
\epsfig{file=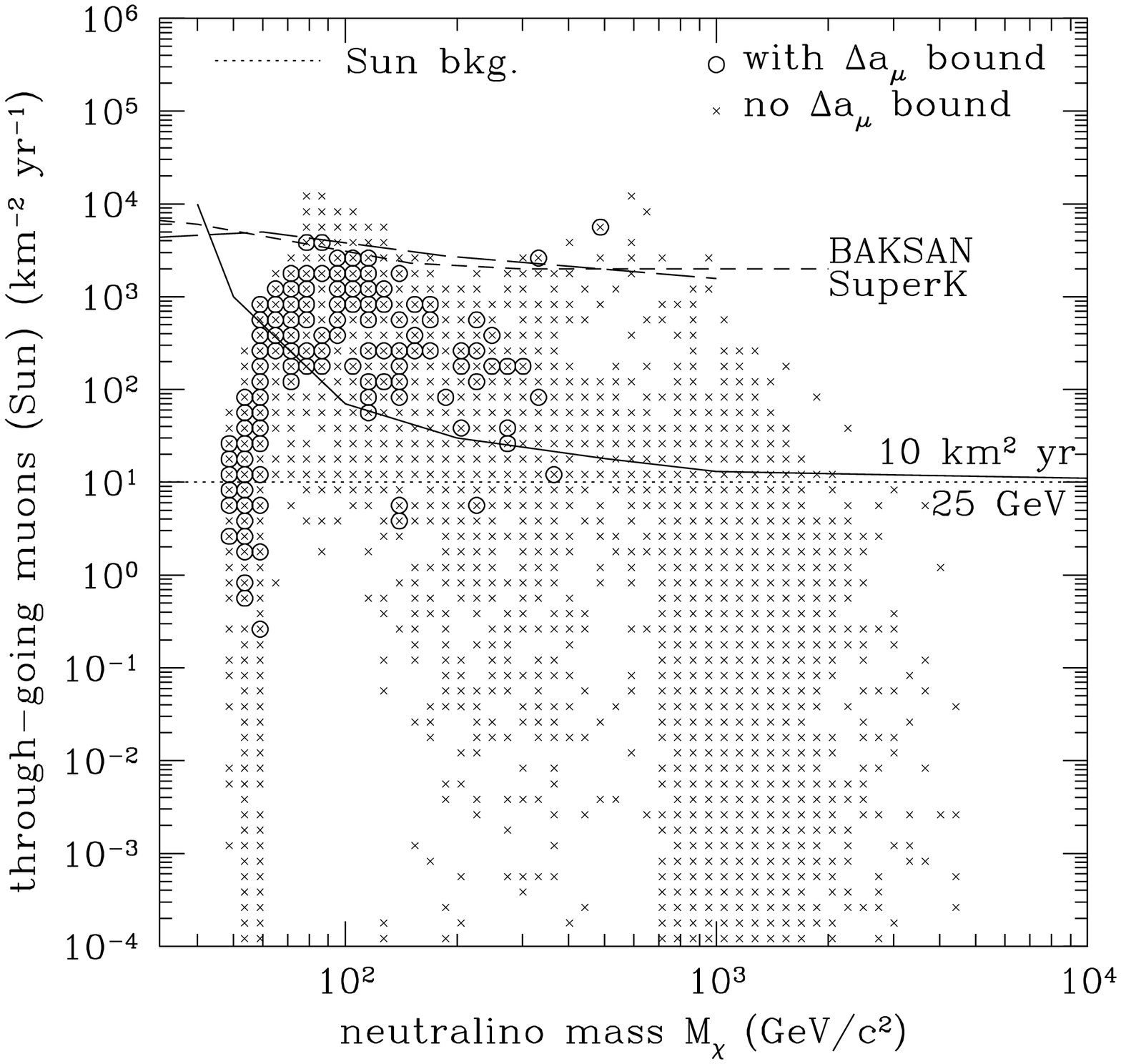,width=0.415\textwidth}}
\centerline{
\epsfig{file=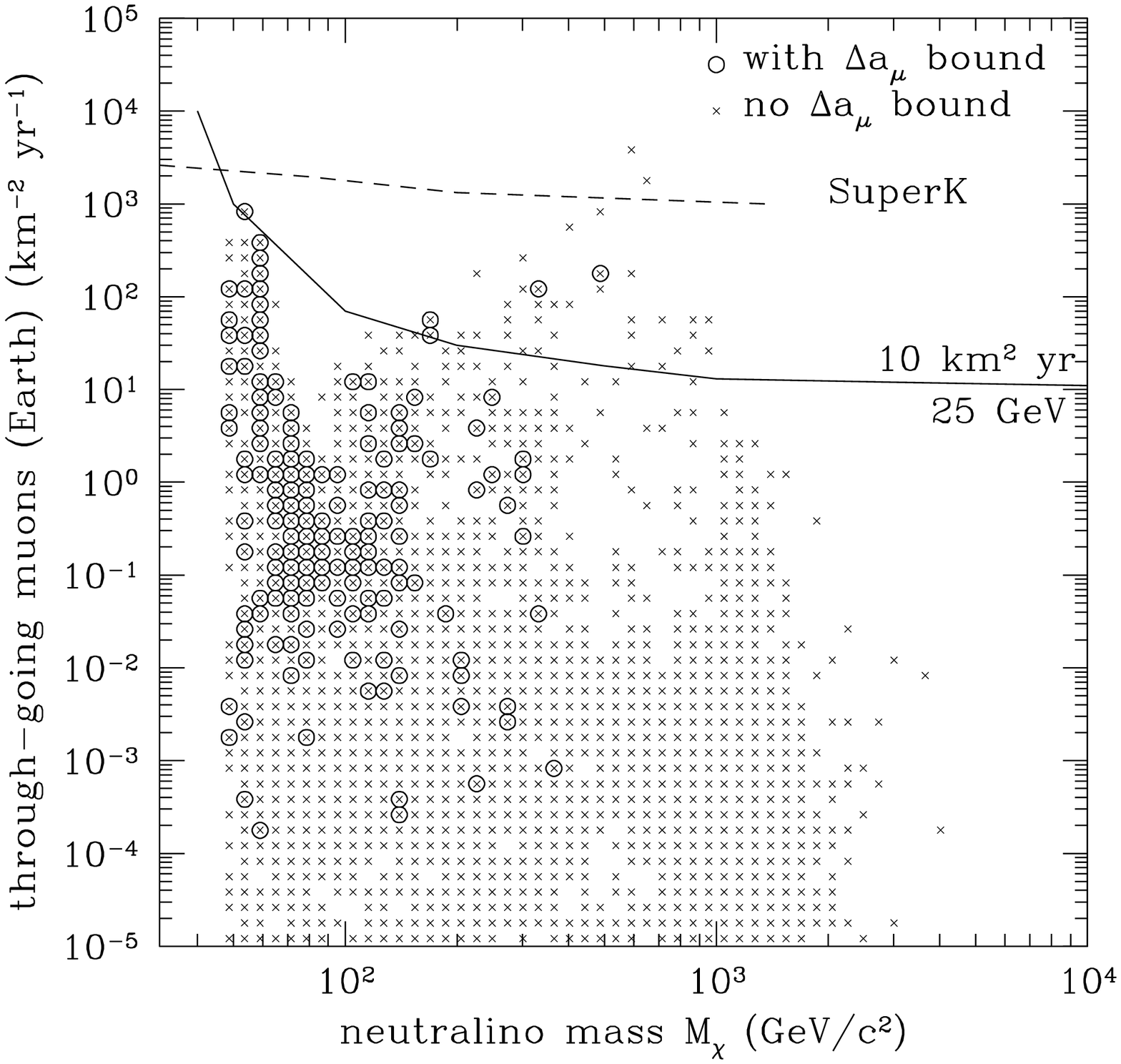,width=0.415\textwidth}
\epsfig{file=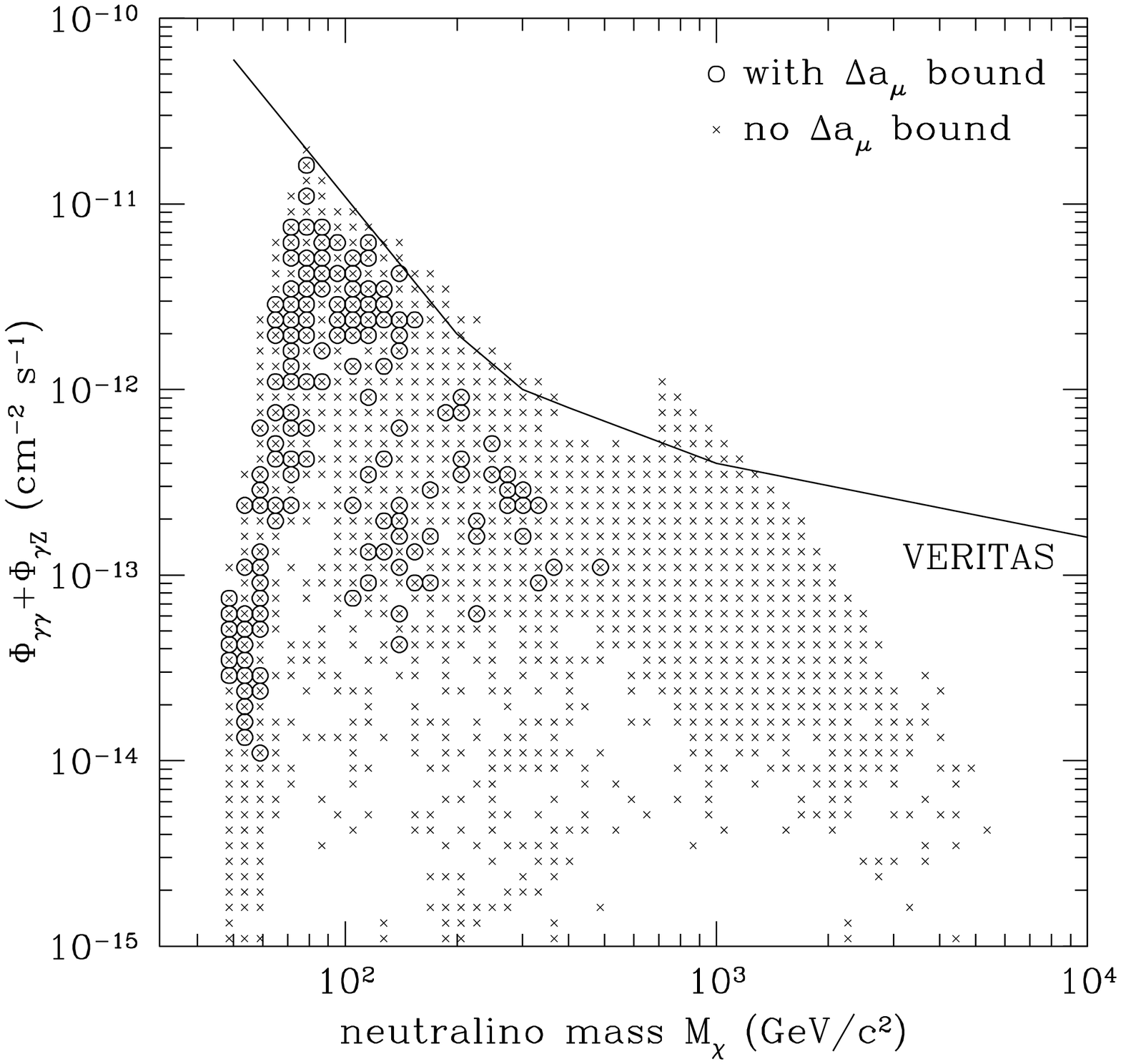,width=0.415\textwidth}}
\caption{Astrophysical detectability of SUSY models.  In all plots, small
crosses indicate cosmologically interesting models, and crossed circles
indicate such models that pass the $\dams$ cut.  In the top left we plot the
spin-independent cross section for neutralino-proton scattering, combined with
the CDMS bound and the reach of GENIUS.  In the top right we plot the rate of
through-going muons in a neutrino telescope for the annihilations in the Sun,
with the BAKSAN and SuperKamiokande bounds, and the reach of a km$^2$
telescope. In the bottom left, we plot a similar rate for neutrinos from the
center of the Earth.  In the bottom right we plot the intensity of the
gamma ray lines in the direction of the galactic center, with the future reach
of the VERITAS experiment \protect\cite{gammas}.}
\label{fig:detect}
\end{figure*}

One of the most promising astrophysical techniques for detecting neutralino
dark matter is the so-called direct detection program.  Neutralinos in the
galactic halo are constantly passing through the Earth, and may be detectable
with sensitive underground instruments such as CDMS \cite{cdms} and DAMA
\cite{dama}. The neutralino--nucleon elastic scattering cross section is
correlated with $\dams$ \cite{drees}. In the top left panel of
Fig.~\ref{fig:detect}, we plot the spin-independent neutralino-proton
scattering cross section.  The constraint due to $\dams$ is intriguing, as it
raises the minimum cross section by many orders of magnitude, to around
$10^{-9}$ pb.  This is very interesting in that it places a bound that is
conceivably detectable in future experiments, such as GENIUS \cite{genius}.

Another possible method to detect neutralino dark matter is neutrino
telescopes, such as at Lake Baikal \cite{baikal}, Super-Kamiokande
\cite{superk}, in the Mediterranean \cite{antares}, and the south pole
\cite{amanda}.  Neutralinos in the galactic halo undergo scatterings into bound
orbits around the Earth and Sun, and subsequently sink to the centers of these
bodies.  The resulting enhanced density can produce a detectable annihilation
signal in neutrinos at GeV and higher energies.  The detectability of this
signal is strongly correlated with the capture rate, which in turn is strongly
correlated with the neutralino-nucleon cross sections discussed in the previous
paragraph.  Thus, there is a much more promising lower bound on the neutrino
flux from the Sun, though the flux from the Earth can still be quite small.  To
illustrate, we plot the rate of neutrino-induced through-going muons from the
Sun, along with the unsubtractable background, in the top right panel of
Fig.~\ref{fig:detect}.  We see that the $\dams$ bound removes most undetectable
models, though there remain some such models at low neutralino masses, as they
suffer from threshold effects \cite{neutrate}.  The flux of neutrinos from the
Earth is plotted in the bottom right panel.

Gamma ray experiments such as atmospheric \v Cerenkov telescopes (ACTs)
can in principle detect the annihilation lines of dark matter neutralinos in
the galactic halo directly either to two photons, or to a photon and a $Z$
boson.  In removing the high-mass models, the reach of ACTs is limited, as they
tend to have thresholds above 100 GeV \cite{bub}.  Furthermore, we see that
applying the $\dams$ bound (bottom right panel of Fig.~\ref{fig:detect}) does
not greatly increase the lower bound of gamma ray flux.

In this Letter we have discussed some implications of the recent measurement of
the anomalous magnetic moment of the muon \cite{DATA}.  In particular, we have
shown that in taking the measurement at face value, the constraints placed on
the supersymmetric parameter space significantly improve the prospects for
direct detection experiments seeking to measure the infrequent scatterings of
galactic halo neutralinos and neutrino telescopes seeking the annihilation
signals from the centers of the Earth and Sun.  Searches for the monochromatic
gamma rays from neutralino annihilation towards the galactic center are not
helped or hindered much by this result.

EB thanks A.~J.~Baltz for forwarding the press release and preprint announcing
the experimental result discussed in this Letter.


\end{document}